
\input harvmac
\input epsf

\newcount\figno
\figno=0
\def\fig#1#2#3{
\par\begingroup\parindent=0pt\leftskip=1cm\rightskip=1cm\parindent=0pt
\baselineskip=11pt
\global\advance\figno by 1
\midinsert
\epsfxsize=#3
\centerline{\epsfbox{#2}}
\vskip 12pt
{\bf Figure \the\figno:} #1\par
\endinsert\endgroup\par
}
\def\figlabel#1{\xdef#1{\the\figno}}

\def\N{{\cal N}}
\def\Q{{\cal Q}}
\def\th{\theta}
\def\d{\delta}
\def\e{\epsilon}
\def\vph{\varphi}
\def\sm#1{\sum^{N}_{{#1}=1}}
\def\r{{\bf r}}
\def\D{{\bf D}}
\def\B{{\bf B}}
\def\E{{\bf E}}
\def\S{{\bf S}}
\def\Z{{\bf Z}}
\def\tr{{\rm tr}}
\def\diag{{\rm diag}}
\def\hf{{1\over 2}}
\def\qu{{1\over 4}}

\def\np#1#2#3{{ Nucl. Phys.} {\bf B#1}, #2 (19#3)}

\def\npps#1#2#3{{ Nucl. Phys. Proc. Suppl.} {#1B}, #2 (19#3)}
\def\pln#1#2#3{Phys. Lett. {\bf B#1}, #2 (19#3)}
\def\plo#1#2#3{{ Phys. Lett.} {\bf #1B}, #2 (19#3)}
\def\pr#1#2#3{{ Phys. Rev.} {\bf D#1}, #2 (19#3)}
\def\prl#1#2#3{{ Phys. Rev. Lett.} {\bf #1}, #2 (19#3) }

\def\yz#1#2#3{{ Yad. Fiz.} {\bf #1}, #2 (19#3)}
\def\sjnp#1#2#3{{ Sov. J. Nucl. Phys.} {\bf #1}, #2 (19#3)}
\def\hpt#1{{\tt hep-th/#1}}

\lref\HHS{
K. Hashimoto, H. Hata, and N. Sasakura,
``3-String Junction and BPS Saturated Solutions 
in $SU(3)$ Supersymmetric Yang-Mills Theory'',
\hpt{9803127}.
}
\lref\Bogo{
E. B. Bogomolny,
``Stability of Classical Solutions'',
\yz{24}{861-870}{76},
\sjnp{35}{449-454}{76}.
}
\lref\PS{
M. K. Prasad and C. M. Sommerfield,
``Exact classical solution for the 't Hooft monopole 
and the Julia-Zee dyon'',
\prl{35}{760-762}{75}.
}
\lref\MO{
C. Montonen and D. Olive,
``Magnetic Monopoles as Gauge Particles?'',
\plo{72}{117-120}{77}.
}
\lref\Osb{
H. Osborn,
``Topological Charges for $N=4$ Supersymmetric Gauge Theories 
and Monopoles of Spin 1'',
\plo{83}{321-326}{79}.
}
\lref\FH{
C. Fraser and T. J. Hollowood,
``Semiclassical Quantization in $N=4$ Supersymmetric Yang-Mills Theory 
and Duality'',
\pln{402}{106-112}{97},
\hpt{9704011}.
}
\lref\WB{
D. Wilkinson and F. A. Bais,
``Exact $SU(N)$ monopole solutions with spherical symmetry'',
\pr{19}{2410-2415}{79}.
}
\lref\bound{
A. Sen,
``Dyon-Monopole Bound States, Self-dual Harmonic Forms on the
Multi-Monopole Moduli Space, and $SL(2,\Z)$ Invariance in String Theory'',
\pln{329}{217-221}{94}, \hpt{9402032}.
}
\lref\Sch{
J. H. Schwarz,
``Lectures on Superstring and M theory Dualities'',
\npps{55}{1-32}{97}, \hpt{9607201}.
}
\lref\net{
A. Sen,
``String Network'',
\pln{329}{217-221}{94}, \hpt{9711130}.
}
\lref\BI{
K. Dasgupta and S. Mukhi,
``BPS Nature of 3-String Junctions'',
\hpt{9711094}\semi
S.-J. Rey and J.-T. Yee,
``BPS Dynamics of Triple $(p,q)$ String Junction'',
\hpt{9711202}.
}
\lref\M{
K. Krogh and S. Lee,
``String Network from M-Theory'',
\np{516}{241-254}{98}, \hpt{9712050}\semi
Y. Matsuo and K. Okuyama,
``BPS Condition of String Junction from M theory'',
\hpt{9712070}\semi
I. Kishimoto and N. Sasakura,
``M-theory description of BPS string in 7-brane background'',
\hpt{9712180}.
}
\lref\Berg{
O. Bergman,
``Three-Pronged Strings and 1/4 BPS States in $\N=4$ Super-Yang-Mills
Theory'',
\hpt{9712211}.
}
\lref\table{
I. S. Gradshteyn and I. M. Ryzhik; A. Jeffrey ed., 
``Table of Integrals, Series, and Products (5th ed.)'',
(Academic Press, 1994).
}
\lref\hhs{
K. Hashimoto, H. Hata, and N. Sasakura, 
to appear. 
}

\Title{                                \vbox{\hbox{UT-811}
                                             \hbox{\tt hep-th/9804139}} }
{\vbox{\centerline{
                String Network and 1/4 BPS States 
}
\vskip .1in
\centerline{
      in $\N=4$ $SU(N)$ Supersymmetric Yang-Mills Theory
}}}

\vskip .2in

\centerline{
                  Teruhiko Kawano and Kazumi Okuyama
}

\vskip .2in 

\centerline{
               Department of Physics, University of Tokyo
}
\centerline{
                   Hongo, Tokyo 113-0033, Japan
}
\centerline{\tt
              kawano@hep-th.phys.s.u-tokyo.ac.jp
}

\vskip -.1in
\centerline{\tt
              okuyama@hep-th.phys.s.u-tokyo.ac.jp
}

\vskip 3cm

We construct the classical configurations of BPS states 
with $1/4$ unbroken supersymmetries
in four-dimensional $\N=4$ $SU(N+1)$ supersymmetric 
Yang-Mills theory, and discuss that
these configurations correspond to string networks 
connecting $(N+1)$ D3-branes in Type IIB string theory.

\Date{April, 1998}

\newsec{Introduction} 
The four-dimensional $\N=4$ supersymmetric Yang-Mills
theory (SYM) was conjectured to be invariant 
under $SL(2,\Z)$ duality, which is the generalization of 
the Montonen-Olive duality \MO\ \Osb.  
$SL(2,\Z)$ duality predicts the existence of 
the dyon bound states with relatively prime magnetic and 
electric charges as BPS states with half unbroken supersymmetries 
in $\N=4$ SYM. This prediction was checked for the two-monopole sector
\bound. In contrast to the $\hf$ BPS states in the $\N=4$ SYM, little is
known about BPS states with $1/4$ unbroken supersymmetries, 
which we will call `{\it Yon}' solutions. 
It sounds like dyon, and the letter `Y' looks like a three-string junction 
which is a fundamental element of a string network. We will briefly explain 
the connection between the $\qu$ BPS state and the string junction soon below.

In string theory, the $d=4$ $\N=4$ SYM appears as the effective field theory
on D3-branes. In the D-brane picture, $SL(2,\Z)$ duality of 
the $d=4$ $\N=4$ SYM is that of the type IIB string theory. 
The $(p,q)$ dyon in the $\N=4$ SYM corresponds to a $(p,q)$ string 
stretched between two D3-branes.
In \Berg, it was argued that the Yon solution in the $SU(3)$ SYM 
corresponds to a three-string junction \refs{\Sch\net\BI{--}\M}
which connects three D3-branes.

Recently, a classical Yon solution was constructed for 
the case of $SU(3)$ gauge group in \HHS.  
In this paper, we generalize the result of \HHS\ to the gauge
group $SU(N+1)$. We construct the spherically symmetric solution of 
the $\qu$ BPS equation of the $d=4$ $\N=4$ $SU(N+1)$ SYM. We discuss that 
these solutions can be interpreted in the string theory as 
the string networks with $N-1$ junctions connecting $N+1$ D3-branes.

\newsec{$1/4$ BPS Configurations in $\N=4$ $SU(N+1)$ Supersymmetric
Yang-Mills Theory}

We begin with a summary on the strategy adopted by Hashimoto, Hata, and 
Sasakura \HHS\ to find $1/4$ BPS solution in $\N=4$ $SU(3)$ supersymmetric 
Yang-Mills theory. But, here, we will extend the gauge group 
$SU(3)$ to $SU(N+1)$. 

In $\N=4$ $SU(N+1)$ super Yang-Mills theory, we have six real scalar fields, 
a gauge field, and a gaugino. To find our BPS solutions, we turn off the 
gaugino field and the four scalar fields. The two remaining real scalars 
$\tilde X$, $\tilde Y$ are expected to describe strings 
which form a string network on the two-dimensional plane. 
In \FH \HHS, the BPS condition \Bogo \PS\ for 
$1/4$ unbroken supersymmetric solutions was given\foot{Our condition here 
is more general than that in \FH \HHS. See appendix A.} by 
\eqn\BPS{\eqalign{
&\left(\matrix{\D\tilde X \cr \D\tilde Y}\right)
=\left(\matrix{\cos\th & -\sin\th \cr \sin\th & \cos\th \cr}\right)
\left(\matrix{\E \cr \B}\right),
\cr
&\left(\matrix{D_0\tilde X \cr D_0\tilde Y}\right)
=i\big[ \tilde X, \tilde Y \big]\left(\matrix{-\sin\th \cr \cos\th}\right),
\cr
&\D\cdot\E
=i\big[ \tilde X, D_0\tilde X \big]+i\big[ \tilde Y, D_0\tilde Y \big].
\cr}
}
By the rotation on the two-dimensional plane $(\tilde X, \tilde Y)$; 
\eqn\rot{
\left(\matrix{\tilde X \cr \tilde Y}\right)
=\left(\matrix{\cos\th & -\sin\th \cr \sin\th & \cos\th \cr}\right)
\left(\matrix{X \cr Y}\right),
}
we rewrite \BPS\ into
\eqn\bps{\eqalign{
&\left(\matrix{\D X \cr \D Y}\right)
=\left(\matrix{\E \cr \B}\right),
\cr
&\left(\matrix{D_0X \cr D_0Y}\right)
=i\big[ X, Y \big]\left(\matrix{ 0 \cr 1 }\right),
\cr
&\D\cdot\E
=i\big[ X, D_0X \big]+i\big[ Y, D_0Y \big].
\cr}
}

Since we are interested in static solutions, we will drop the time-dependence 
of all the fields. Then, putting $X=-A_0$, we can see that the two following  
equations remains to be solved: 
\eqn\Mpole{
\D Y = \B,
}
\eqn\Yon{
\D\cdot\D X = -i\left[Y, \left[Y, X\right]\right].
}
The first equation \Mpole\ is the well-known BPS condition \Bogo \PS\ 
for monopole solutions. As the monopole solution, we adopt spherically 
symmetric solutions given in \WB, as the authors of \HHS\ have done for 
$SU(3)$ gauge group.
The solution is given by 
\eqn\monopole{
{\bf A}(\r)=\left({\bf M}(\r)-{\bf T}\right)\times{\r \over r^2},
}
where ${\bf T}$ is the maximal $SU(2)$ embedding in $SU(N+1)$ with 
$T_3={\rm diag}({1\over2}N, {1\over2}N-1, \cdots, -{1\over2}N+1, -{1\over2}N)$.
On the $z$-axis, $M$ and $Y$ have the following forms: 
\eqn\ansatz{\eqalign{
&M_{+}=M_1+iM_2=\sm{k} a_k(r) E_k,
\cr
&Y=\hf\sm{k} \phi_k(r) H_k,
\cr}
}
which will be determined so as to satisfy the equation \Mpole\ on the $z$-axis
\eqn\mnp{\eqalign{
r^2Y'&={1\over2}\left[M_+,M_-\right]-T_3,
\cr
M_{\pm}'&=\mp\left[M_{\pm},Y\right].
\cr}
}
Here $H_k$ $(k=1,\cdots,N)$ is the generator in the Cartan subalgebra 
of $SU(N+1)$. $E_k$ $(k=1,\cdots,N)$ is the generator which corresponds to 
a simple root. As a concrete representation, 
$(H_k)_{a,b}=( \d_{a,k}\d_{b,k}-\d_{a,k+1}\d_{b,k+1} )$ and 
$(E_k)_{a,b}=( \d_{a,k}\d_{b,k+1} )$; $a,b=1, \cdots, N+1$. 
Then, $[E_k, E^{\dag}_l]=\d_{k,l}H_k$ and $[H_k, E_l]=C_{l,k}E_{l}$, 
where $C_{l,k}=2\d_{l,k}-\d_{l,k+1}-\d_{l,k-1}$ is the Cartan matrix 
of $SU(N+1)$.

In particular, the functions $a_k(r)$ and $\phi_k(r)$ are expressed 
by $N$ functions $Q_k(r)$ $(k=1, \cdots, N)$;
\eqn\apq{\eqalign{
&a_k(r)={r \over Q_k}\left(k\bar k Q_{k-1}Q_{k+1}\right)^{{1\over2}},
\cr
&\phi_k(r)=-{d\ln Q_k \over dr}+{{k \bar k} \over r},
\cr}
}
with $\bar k=N+1-k$. Here we assume $Q_0=Q_{N+1}=1$.
These functions $Q_k(r)$ satisfy the following differential equation: 
\eqn\diff{
Q'_kQ'_k-Q_kQ''_k = k \bar k Q_{k-1} Q_{k+1},
}
for $k=1,2, \cdots, N$. Here the prime denotes the differentiation with 
respect to the radial coordinate $r$.

Although the general solution $Q_k$ for the equation \diff\ are given in \WB, 
it was difficult for the authors of this paper to solve the second equation 
\Yon\ with the general solution $Q_k$. Instead of it, we find a solution 
\eqn\spsol{
Q_k(r)=\left({\sinh cr \over c}\right)^{k\bar k},
}
which corresponds to the solution in \WB\ with a particular value 
of $N$ parameters. We also notice that, for $N+1=3$, ours \spsol\ is 
the same solution as that adopted in \HHS. 

Therefore, inserting the solution \spsol\ into \Yon\ on the $z$-axis;
\eqn\yon{
\left(rX\right)''-{1 \over 2r^2}\left(\big[ M_{+} , \big[ M_{-} , rX \big]\big]
+\big[ M_{-} , \big[ M_{+} , rX \big]\big]\right)
=-i\big[ Y , \big[Y , rX \big]\big],
}
and taking an ansatz for $rX(r)$ as $rX(r)=\sm{k} f_k H_k$, we obtain 
\eqn\keyeq{
f''_k-{c^2 k \bar k \over \sinh^2rc}\left( 2f_k-f_{k+1}-f_{k-1} \right)=0,
}
with $f_0=f_{N+1}=0$. In order to diagonalize this equation \keyeq, 
we rewrite the functions $f_k$ with  
$N$ new functions $\vph_s$ $(s=1, \cdots, N)$ by 
$f_k=\sm{s} v^{(s)}_k \vph_s$, 
where $v^{(s)}$ is the eigenvector of
the matrix $\tilde{C}=(\tilde{C}_{k,l}):=(k\bar k C_{k,l})$
with the eigenvalue $s(s+1)$.
The generating function $F^{(s)}(t)$ of $v^{(s)}$ is given by
\eqn\gen{\eqalign{
F^{(s)}(t) &:=\sum_{k=1}^{N}v^{(s)}_kt^{k-1} \cr
&=\sum_{n=0}^{N-s}\left(\matrix{N+s+1 \cr N-s-n} \right)
\left(\matrix{n+s \cr s}\right)(t-1)^{n+s-1},
\cr}
} 
or the explicit form of $v^{(s)}_k$ is
\eqn\trfv{
v^{(s)}_k=\sum^{N-s}_{n=n_0}(-1)^{n+s-k}
\left(\matrix{N+s+1 \cr N-s-n}\right)
\left(\matrix{n+s \cr s}\right)
\left(\matrix{n+s-1 \cr k-1}\right),
}
where $n_0={\rm max}\{0,k-s\}$.
We can see that the equation \keyeq\ turns into 
\eqn\Lgdreq{
{d^2 \over dx^2} \vph_s(r) = {1 \over \sinh^2x}s(s+1)\vph_s(r)
}
with $x=rc$.

The equation \Lgdreq\ can be seen to be the Legendre differential equation, 
if we introduce a new variable $\xi$ given by $\xi=\coth x$ and rewrite 
\Lgdreq\ with $\xi$. 

Imposing the boundary condition on $\vph_s(r)$ by 
$\vph_s(r=0)=\vph'_s(r=0)=0$, we can see that 
the solution $\vph_s(r)$ is given 
by $\vph_s(r)=c_s\Q_s(\xi)$ with $c_s$ any proportional constant, 
where $\Q_s(\xi)$ is Legendre function of the second kind \table;
\eqn\Lgdrfn{
\Q_s(\xi)={1\over2^{s+1}}\int^{1}_{-1}(1-t^2)^s(\xi-t)^{-s-1}dt.
}

\newsec{Discussion}

\fig{ 
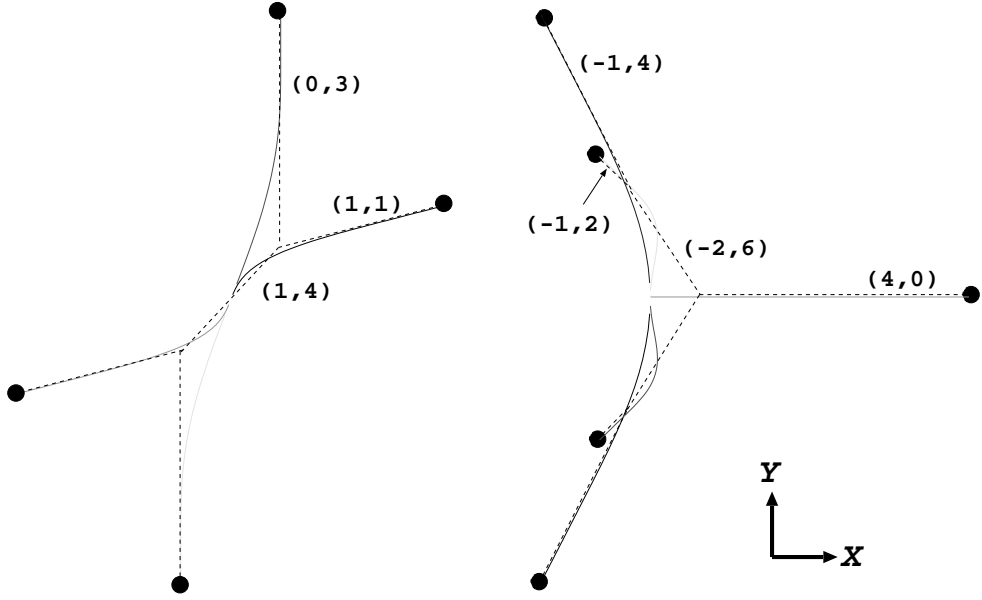
The curved lines in the left(right) figure represent the configuration of Yon in
$SU(4)$($SU(5)$) gauge theory.  The dashed straight lines are the corresponding
string network, and the row vectors are the IIB- and D-string charges
of each segment. The dots represent the asymptotic positions of D3-branes.   
}{su45-mod2.eps}{13 truecm}
\figlabel\su

We have constructed the 1/4 BPS configurations in $\N=4$ $SU(N+1)$ SYM 
in the previous section. The configurations of two real scalars $X$, $Y$ 
are 
\eqn\solution{\eqalign{
X&= {1 \over r}\sm{s}\sm{k} c_s\Q_s(\coth cr)\,v^{(s)}_kH_k,
\cr
Y&= -{\Q_1(\coth cr) \over r} T_3,
\cr}
}
where $c_s$ is an integration constant.
For $N=2$, our solution is identical to that in \HHS. 
For $N=3, 4$, these configurations are drawn in Fig. 1.

From the asymptotic behaviour of $X$ and $Y$ in the limit
$r\rightarrow\infty$,
\eqn\asympt{\eqalign{
X &\sim \diag\left(x_1, \cdots, x_{N+1}\right)
+{1\over2r}\diag\left(e_1, \cdots, e_{N+1}\right),
\cr
Y &\sim \diag\left(y_1, \cdots, y_{N+1}\right)
+{1\over2r}\diag\left(g_1, \cdots, g_{N+1}\right),
\cr}
}
we can see the positions $(x_i, y_i)$ of $D3$-branes and the electric 
and the magnetic charges $(e_i, g_i)$ of strings with their ends on 
the branes \HHS, which are
\eqn\charge{\eqalign{
&(x_i,y_i)=\left(|c|\sum_{s=1}^N\big({\rm sgn}(c)\big)^{s-1}(v_i^{(s)}-v_{i-1}^{(s)})c_s~,
 ~-{|c|\over 2}\bigl\{N-2(i-1)\bigr\} \right),   \cr
&(e_i,g_i)=\left(-2\sum_{s=1}^N\big({\rm sgn}(c)\big)^{s-1}(v_i^{(s)}-v_{i-1}^{(s)})c_s
\sum_{n=1}^s{1\over n}~,~ N-2(i-1)\right).
\cr}}

Our solutions are not general one with arbitrary values of the parameters 
of the monopole solutions \WB. It was hard for us to obtain such general 
solutions, even for $SU(3)$ case. But it should be interesting to seek them, 
when we consider the correspondence between them and the string network. 
Also, it is unclear whether there are any networks with closed loops among 
our solutions, although it is unlikely that our solutions include 
such loops. Then we are led to ask how we can obtain a Yon solution in 
SYM which corresponds to such a network with closed loops. 
We hope to return to these problems in the near future.

\bigskip
\centerline{{\bf Acknowledgements}}
We are grateful to Kazuo Hosomichi and Masatoshi Sato for discussions.
K.O. is supported in part by JSPS Research Fellowships for Young
Scientists.

\bigskip
\noindent
{\bf Note added}: After submitting our paper to the e-print archive,
we were informed that Hashimoto, Hata, and Sasakura have also obtained
the Yon solutions\hhs.   

\appendix{A}{The 1/4 BPS condition in $\N=4$ Supersymmetric Yang-Mills Theory} 

The bosonic part of action of four-dimensional $\N=4$ $SU(N)$ supersymmetric 
Yang-Mills theory 
is given by 
\eqn\action{
S=\int d^4x\, \tr\left[-{1\over4}F_{\mu\nu}F^{\mu\nu}
-{1\over2}D_{\mu}X^ID^{\mu}X^I+{1\over4}\big[X^I, X^J\big]^2\right], 
}
with $\mu, \nu=0,1,2,3$ and $I, J=1,\cdots, 6$, 
where $F_{\mu\nu}=\partial_{\mu}A_{\nu}-\partial_{\nu}A_{\mu}-
i[A_{\mu},A_{\nu}]$ and $D_{\mu}X=\partial_{\mu}X-i[A_{\mu},X]$.
From this action, we obtain the Gauss law
\eqn\gauss{
\D\cdot\E=i\big[ X^I , D_0X^I \big],
}
where $\E$ is the electric field, defined by $E_i=F_{0i}$.

The energy of this system is given by 
\eqn\energy{
E=\int d^3x\, {1\over2} \tr 
\bigg[ \E^2+\B^2+(D_0X^I)^2+(\D X^I)^2-{1\over2}\big[X^I,X^J\big]^2 \bigg],
}
where $\B$ is the magnetic field, defined by $B_i={1\over2}\e_{ijk}F_{jk}$.

Henceforth, we keep the gauge field $A_{\mu}$ and only the two scalar fields 
$X^1=X$, $X^2=Y$ nonvanishing 
and, using the Gauss law \gauss, rewrite the energy 
\energy\ with a parameter $\th$ into
\eqn\sqE{\eqalign{
E&=\int d^3x\, {1\over2} \tr\bigg[
\left( \cos\th \E - \sin\th \B - \D X \right)^2
+\left( \sin\th \E + \cos\th \B - \D Y \right)^2
\cr
&\qquad \qquad \qquad \ +\left(D_0X+i\sin\th \big[X,Y\big]\right)^2
+\left(D_0Y-i\cos\th \big[X,Y\big]\right)^2
\bigg]
\cr
& \qquad \qquad \qquad \ +\left(Q_X+M_Y\right)\cos\th+\left(Q_Y-M_X\right)\sin\th
\cr
&\geq \left(Q_X+M_Y\right)\cos\th+\left(Q_Y-M_X\right)\sin\th
\cr}
}
where $Q_X=\int d\S\,\tr[\,\E X ]$ and  $M_X=\int d\S\,\tr[\,\B X ]$, 
and similarly for $Q_Y$ and $M_Y$.

Since the energy $E$ in the left hand side of \sqE\ is independent of 
the parameter $\th$, the inequality should be hold for any value of the 
parameter $\th$. Thus, in order for the inequality to hold in such a way, 
the energy $E$ must satisfy an inequality $E \geq M$ with 
$M=[(Q_X+M_Y)^2+(Q_Y-M_X)^2]^{1/2}$.

When the value of the parameter $\th$ is such that 
$\tan\th=(Q_Y-M_X)/(Q_X+M_Y)$, the energy $E$ saturates the bound $M$, 
if the configuration of the fields satisfies the BPS condition \BPS, 
as you can see from the right hand side of the equation \sqE.
Also, note that, since we use the Gauss law \gauss\ to obtain \sqE, 
the configuration should satisfy the Gauss law \gauss, too.

\listrefs

\end